\documentclass[11pt,a4paper,amsfonts,amssymb,amsmath,axodraw]{article}
\usepackage {axodraw}
\usepackage{amsmath,amsxtra,amssymb,latexsym, amscd}
\usepackage[mathscr]{eucal}

\makeindex
\begin{document}
\parindent=1.05cm 
\setlength{\baselineskip}{12truept} \setcounter{page}{1}
\makeatletter
\renewcommand{\@evenhead}{\@oddhead}
\renewcommand{\@oddfoot}{}
\renewcommand{\@evenfoot}{\@oddfoot}
\renewcommand{\thesection}{\arabic{section}.}
\renewcommand{\theequation}{\thesection\arabic{equation}}
\@addtoreset{equation}{section}
\title{}
\date{} 
\maketitle \pagestyle{plain} \pagestyle{myheadings}
\markboth{\footnotesize {\textbf{Nguyen Duc Minh}}}
{\footnotesize{\textbf{Path Integral Quantization Of
Self-Interacting Scalar Field With Higher Derivatives}}}
\vskip-4cm 

\begin{center}
\vskip0.5cm 
{\large \bf PATH INTEGRAL QUANTIZATION OF SELF INTERACTING SCALAR
FIELD WITH HIGHER DERIVATIVES}
\end{center}
\vskip0.1cm 
\centerline{\textbf{Nguyen Duc Minh}\footnote{email:
ndthe@physics.org }}
\vskip0.1cm 
\centerline{\it Department of Physics, College of Science,Vietnam
National University, Hanoi}
\begin{quote}
{\small \noindent {\bf Abstract:} \it Scalar field systems
containing higher derivatives are studied and quantized by
Hamiltonian path integral formalism. A new point to previous
quantization methods is that field functions and their time
derivatives are considered as independent canonical variables.
Consequently, generating functional, explicit expressions of
propagators and Feynman diagrams in $\phi^3$ theory are found.}
\end{quote}

\leftskip0cm 
\vskip0.4cm 

\section{INTRODUCTION}
\vskip0.1cm 
\hspace{1cm} Field systems containing derivatives of order higher
than first have more and more important roles with the advent of
super-symmetry  and string theories \cite{ref1}. However, up to now
path integral quantization method is almost restricted to fields
with first derivatives \cite{ref2,ref3,ref4}.
\vspace{6pt}\\
\indent The purpose of this paper is to apply the new ideal
``\emph{velocities have to be taken as independent canonical
variables}'' \cite{ref5} to extending the method to self-interacting
scalar field containing higher derivatives.\vspace{6pt}\\
\indent The paper is organized as follows: Section II presents the
application of this quantization method to quantizing free scalar
field with higher derivatives. Section III is devoted to studying
the Feynman diagrams of self-interacting scalar field. Section IV is
for the drawn conclusion.
\vskip0.4cm 

\section{FREE SCALAR FIELD}
 \hspace{1cm} Let us consider Lagrangian density for a free
scalar field, containing second order derivatives
\begin{equation} \label{1} \tag{1}
L = \frac{1}{2}\left( {\partial _\mu  \phi \,\partial ^\mu  \phi  -
m^2\, \phi ^2 } \right) + \frac{1}{{2\Lambda ^2 }}\,\square\,\phi
\,\square\,\phi,
\end{equation}
where $\square$ is D'Alamber operator
$(\square=\partial_\mu\,\partial ^\mu=\frac{\partial^2}{\partial
t^2}-\triangle)$, $\triangle$ is Laplacian and $\Lambda$ is a
parameter with dimension of mass. It will give a term with $k^4$ in
the denominator of the corresponding Feynman propagator. This
renders a finite result for some diagrams and, consequently, it may
permit the introduction of convenient counter-terms to absorb the
infinities which appears when the limit
$\Lambda$ is taken.\vspace{6pt}\\
\indent The canonical momenta, conjugating to $\phi$ and
$\dot{\phi}$, are respectively
\begin{equation}\label{2}\tag{2}
\pi=\dot{\phi}-\frac{1}{\Lambda^2}\,\square\,\dot{\phi}\,; \qquad
\qquad s=\frac{1}{\Lambda^2}\,\square\,\phi.
\end{equation}
Now, there are no constraints involved. To implement the path
integral quantization of this field we have to pay attention to the
fact that $\dot\phi$ is now an independent canonical variable and,
consequently, it has to be functionally integrated. Thus, the
canonical Hamiltonian density becomes
\begin{equation}\label{3}\tag{3}
\begin{split}
\mathscr{H}_c&=\pi\dot{\phi}+s\ddot{\phi}-L\\
&=\pi X + \frac{1}{2}\Lambda ^2\,s^2  + s\,\nabla ^2 \phi  -
\frac{1}{2}X^2  + \frac{1}{2}\left( {\nabla \phi } \right)^2  +
\frac{1}{2}m^2\,\phi ^2,
\end{split}
\end{equation}
where to avoid mistakes we have denoted the independent coordinate $\dot\phi$ by X.\vspace{6pt}\\
\indent The corresponding generating functional is
\begin{equation}\label{4}\tag{4}
  \begin{aligned}
  Z\left[{J,K}\right]\,&=&N\int{\left[{d\phi}\right]\left[{ds}
  \right]\left[{d\pi}\right]\left[{dX}\right]\textrm{exp}\left\{
  {i\int {d^4 x\left[ {\pi \dot \phi + s\dot X - \pi X -{} }
  \right.}
  } \right. }\\
  \,&&\left.\left.{}-s\nabla^2 \phi
  +\frac{1}{2}X^2-\frac{1}{2}\left(\nabla \phi\right)^2
  -\frac{1}{2}m^2\,\phi^2+J\phi+KX \right]\right\}.
  \end{aligned}
\end{equation}
In this case, integrations over $\pi$ and X are immediately
calculated by using delta function properties and 4-dimensional
Gaussian integral. Integration over $\phi$ is calculated by putting
$\phi=\phi_c + \psi$, in which, $\phi_c$ is determined by field
equation for extended Lagrangian, it means, satisfying
\begin{equation}\label{5}\tag{5}
\left({m^2+\square-\frac{1}{2\Lambda^2}\square\square}\right)=J-\dot{K}.
\end{equation}
\indent The result is
\begin{equation}\label{6}\tag{6}
  \begin{aligned}
  &Z\left[{J,K}\right]=N_1 \textrm{exp}\left\{\frac{i}{2}\int{d^4
  x\left[J\left({x}\right)\frac{1}{\square+m^2-\frac{1}{\Lambda^2}\square\square}\,J\left({x}\right)\right.}\right.\\
  &\phantom{Z\left[{J,K}\right]=N_1
  \textrm{exp}\left\{\frac{i}{2}\int{d^4
  x}\right.}{}-K\left({x}\right)\frac{\partial_0^2}{\square+m^2-\frac{1}{\Lambda^2}\square\square}\,K\left({x}\right)\\
  &\phantom{Z\left[{J,K}\right]=N_1
  \textrm{exp}\left\{\frac{i}{2}\int{d^4
  x}\right.}\left.\left.+2K\left({x}\right)\frac{\partial_0}{\square+m^2-\frac{1}{\Lambda^2}\square\square}\,J\left({x}\right)\right]\right\}.
  \end{aligned}
\end{equation}
\indent The Feynman propagator $\left\langle 0 \right|T\left( {\phi
\left( x \right)\phi \left( {x'} \right)} \right)\left| 0
\right\rangle$ can be directly obtained by the usual expression
\begin{equation}\label{7}\tag{7}
\begin{split}
\left\langle 0 \right|T\left( {\phi \left( x \right)\phi \left( {x'}
\right)} \right)\left| 0
\right\rangle&=\left.\frac{i^{-2}}{Z}\frac{\delta^2Z}{\delta
J\left({x}\right)\delta J\left({x'}\right)}\right|_{J,K=0}\\
&=-\frac{i}{m^2+\square-\frac{1}{\Lambda^2}\square\square}\,\delta^4\left({x-x'}\right).
\end{split}
\end{equation}
\indent Since we have introduced a source for $\dot\phi$, the
following propagators can be obtained
\begin{align}
\left\langle 0 \right|T\left( {\dot\phi \left( x \right)\dot\phi
\left( {x'} \right)} \right)\left| 0
\right\rangle&=\frac{i\,\partial_0^2}{m^2+\square-\frac{1}{\Lambda^2}\square\square}\,\delta^4\left({x-x'}\right), \tag{8}\\
\left\langle 0 \right|T\left( {\phi \left( x \right)\dot\phi \left(
{x'} \right)} \right)\left| 0
\right\rangle&=\frac{-i\,\partial_0}{m^2+\square-\frac{1}{\Lambda^2}\square\square}\,\delta^4\left({x-x'}\right).\tag{9}
\end{align}
\indent Propagator (\ref{7}) is in agreement with the correct
propagator by following the usual canonical procedure \cite{ref6}.
More over, when the limit $\Lambda$ is taken, it has usual form
corresponding to the ordinary free scalar field (containing first
derivatives) we have known before. The above propagators calculated
explicitly is an important step to obtain Feynman diagrams and
propagators of self-interacting scalar field in the next section.
\vskip0.2cm 

\section{SCALAR FIELD IN $\boldsymbol{\phi^3}$
THEORY} \hspace{1cm} Now, we consider $\phi^3$ self-interacting
scalar field by adding an interacting term
$L_{int}=-\frac{g}{6}\phi^3$ to the Lagrangian (\ref{1})
\begin{equation}\label{8}\tag{10}
L =\frac{1}{2}\left({\partial _\mu \phi \,\partial ^\mu  \phi  - m^2
\phi ^2 } \right) + \frac{1}{{2\Lambda ^2 }}\,\square\,\phi\,
\square\,\phi + \frac{g}{6}\phi ^3.
\end{equation}
\indent Since the interacting field $L_{int}$ only depends on
$\phi$ and the final form of the generating functional $Z$ contains
only field configuration $d\phi$ under the integral, the generating
functional $Z\left[{J,K}\right]$ with higher derivatives, in
$\phi^3$ interacting theory, is similar to the ones with first order
derivatives. It means, the re-normalization generating functional
\cite{ref7} $Z\left[{J,K}\right]$ is
\begin{equation}\label{9}\tag{11}
Z\left[ {J,K} \right] = \frac{\textrm{exp}\left[i\int{L_{int}
\left(\frac{1}{i}\frac{\delta}{\delta J}dx\right)}\right]Z_0
\left[{J,K}\right]} {\left.\textrm{exp}\left[i\int{L_{int}
\left(\frac{1}{i}\frac{\delta}{\delta J}dx\right)}\right]Z_0
\left[{J,K}\right]\right|_{J,K=0}}.
\end{equation}
\indent Since  $L_{int}$ also depends only on $\phi$, the formula of
the S matrix still has form
\begin{equation}\label{10}\tag{12}
S=:\textrm{exp}\left[\int{\phi_{int}K\frac{\delta}{\delta
J\left({z}\right)}}\right]:\left.Z\left[{J,K}\right]\right|_{J,K=0},
\end{equation}
where $K=\square+m^2-\frac{1}{\Lambda^2}\square\,\square$.\vspace{6pt}\\
 \indent So that, we can apply LSZ formula to the
interaction between two in-particles and two out-particles. The
scattering amplitude is
\begin{equation}\label{11}\tag{13}
   \begin{aligned}
   \left\langle f\left|S-1\right|i\right\rangle&=&\int{d^4x_1\, d^4x_2 \,d^4x'_1 \,d^4
   x'_2\,
   \textrm{e}^{i\left(k_1x_1+k_2x_2-k'_1x'_1-k'_2x'_2\right)}}\,K \left(x_1 \right)K\left(x_2\right)\times{}\\
   &&{}\times K\left(x'_1\right)K\left(x'_2\right)\left\langle 0 \right|T\left( {\phi (x_1 )\phi \left( {x_2 } \right)\phi \left( {x'_1 } \right)\phi \left( {x'_2 } \right)} \right)\left| 0 \right\rangle
   _C \hspace{6pt},
  \end{aligned}
\end{equation}
where $K\left({x_1}\right)\tau
\left({x_1,y}\right)=-i\,\delta^4\left(x_1-y\right)$.\vspace{6pt} \\
\indent Formula (\ref{11}) is calculated explicitly through 4-point
function (the procedure is the same as in \cite{ref7})
\begin{equation}\label{12}\tag{14}
   \begin{aligned}
   &\left\langle{f\left|S-1\right|i}\right\rangle=\left({-ig}\right)^2\int{d^4y\,d^4z\,\tau\left({y-z}\right)\left[\textrm{e}
   ^{i\left({k_1y+k_2y-k'_1z-k'_2z}\right)}\right.}\\
   &\phantom{\left\langle{f\left|S-1\right|i}\right\rangle=\left({-ig}\right)^2\int{d^4y\,d^4z\,\tau\left({y-z}\right)\left[e\right.}}
   {}+\,\textrm{e}
   ^{i\left({k_1y+k_2z-k'_1y-k'_2z}\right)}\\
   &\phantom{\left\langle{f\left|S-1\right|i}\right\rangle=\left({-ig}\right)^2\int{d^4y\,d^4z\,\tau\left({y-z}\right)\left[e\right.}}
   {}\left.+\,\textrm{e}
   ^{i\left({k_1y+k_2z-k'_1z-k'_2y}\right)}\right]+O\left({g^4}\right),
   \end{aligned}
\end{equation}
where
\begin{equation}\label{13}\tag{15}
\tau\left({x-y}\right)=\int{\frac{d^4k}{\left({2\pi}\right)}\frac{-i}{k^2-m^2+i\varepsilon+\frac{1}{\Lambda^2}k^4}\textrm{e}^{ik\left({x-y}\right)}}.
\end{equation}
\indent Substituting (\ref{13}) for (\ref{12}) and integrating over
$dy$ $dz$, we obtain
\begin{equation}\label{14}\tag{16}
    \begin{aligned}
    & \left\langle f \right|S - 1\left| i \right\rangle  = ig^2 (2\pi )^4 \delta (k_1  + k_2  - k'_1  - k'_2 ) \\
    &\phantom{ \left\langle f \right|S - 1\left| i \right\rangle  = ig^2
    (}{}\times \left[ {\frac{1}{{(k_1  + k_2 )^2  - m^2  + \frac{1}{{\Lambda ^2 }}(k_1  + k_2 )^4 }} } \right. \\
    &\phantom{\left\langle f \right|S - 1\left| i \right\rangle  =
    ig^2(\times[i]}{}+ \frac{1}{{(k_1  - k'_1 )^2  - m^2  + \frac{1}{{\Lambda ^2 }}(k_1  - k'_1 )^4 }} \\
    &\phantom{\left\langle f \right|S - 1\left| i \right\rangle  =
    ig^2(\times[i]}{}\left. {\, + \frac{1}{{(k_1  - k'_2 )^2  - m^2  + \frac{1}{{\Lambda ^2 }}(k_1  - k'_2 )^4 }}} \right] + O(g^4
    ).
    \end{aligned}
\end{equation}
\indent From (\ref{14}), we have the following Feynman rules for
the scattering amplitude \vspace{12pt}\\
\begin{center}
    \begin{tabular}{l c c}
    \hline\hline Diagrammatic representation && Factor in S
    matrix\\
    \hline \\\hspace{34pt} \large{Internal line}& \begin{picture}(60,10)(0,0)
    \ArrowLine(0,3)(60,3) \Text(30,8)[b]{k}\end{picture} &\Large{$\frac{{ - i}}{{k^2  - m^2  + i\varepsilon  + \frac{1}{{\Lambda ^2 }}k^4
    }}$}\vspace{8pt}\\\hspace{34pt}
    \large{External line}&\begin{picture}(60,10)(0,0)\Line(0,3)(60,3)\end{picture}&1 \\
    \hspace{34pt} \large{Vertex}&
        \begin{picture}(60,40)(0,2)
        \ArrowLine(0,-16)(30,6) \ArrowLine(0,28)(30,6)
        \ArrowLine(30,6)(60,6)\Vertex(30,6){2}
        \end{picture}
    &\Large{$\frac{{ - i}}{{k^2  - m^2  + i\varepsilon  + \frac{1}{{\Lambda ^2 }}k^4
    }}$}\\\vspace{4pt}\\
    \hline\hline
    \end{tabular}
\end{center}
\hspace{1cm} In summary, by using above improved path integral
quantization method, Feynman diagrams for self-interacting $\phi^3$
scalar field are found. In general, when interacting term is more
complicated, for example it contains derivatives of $\phi$, Feynman
diagrams will have two more new kinds of vertex, corresponding to
interacting vertices $\dot\phi - \phi$ and $\dot\phi - \dot\phi$.
\vskip0.3cm 

\section{ CONCLUSION }
\hspace{1.05cm} We have studied the improved Hamiltonian path
integral formulation for scalar field with higher derivatives and
also considered the system in $\phi^3$ self-interaction. The new
ideal is that derivatives of field functions are considered as
independent canonical variables. Generating functional and explicit
expressions of propagators are calculated. Feynman diagrams for
$\phi^3$ interacting field are obtained explicitly. Extension of
this result to electrodynamics (interacting with matter), string
theory or gravity theory will be studied latter.

\vskip0.4cm 
\pagebreak
\centerline{\bf ACKNOWLEDGMENT}  
\vskip0.2cm 
The author would like to thank Prof. Nguyen Suan Han for his
suggestions of the problem and many useful comments.

\end{document}